\documentclass[apjl]{emulateapj}
\usepackage{comment}
\usepackage{ifthen}
\usepackage{lineno}

\newcommand{\lc}{light curve}
\newcommand{\lcs}{light curves}
\newcommand{\Lc}{Light curve}
\newcommand{\Lcs}{Light curves}

\newcommand{\band}[1]{\ensuremath{#1}~band}

\newcommand{\kms}{\ensuremath{\rm km\,s^{-1}}}
\newcommand{\ms}{\ensuremath{\rm m\,s^{-1}}}

\newcommand{\gcmc}{\ensuremath{\rm g\,cm^{-3}}}
\newcommand{\ergscmsq}{\ensuremath{\rm erg\,s^{-1}\,cm^{-2}}}

\newcommand{\teff}{\ensuremath{T_{\rm eff}}}
\newcommand{\logg}{\ensuremath{\log{g}}}
\newcommand{\vsini}{\ensuremath{v \sin{i}}}
\newcommand{\feh}{\ensuremath{\rm [Fe/H]}}

\newcommand{\rhk}{\ensuremath{R^{\prime}_{\rm HK}}}
\newcommand{\logrhk}{\ensuremath{\log\rhk}}

\newcommand{\rsun}{\ensuremath{R_\sun}}
\newcommand{\msun}{\ensuremath{M_\sun}}
\newcommand{\lsun}{\ensuremath{L_\sun}}

\newcommand{\rstar}{\ensuremath{R_\star}}
\newcommand{\mstar}{\ensuremath{M_\star}}
\newcommand{\lstar}{\ensuremath{L_\star}}

\newcommand{\teffstar}{\ensuremath{T_{\rm eff\star}}}
\newcommand{\rhostar}{\ensuremath{\rho_\star}}
\newcommand{\loggstar}{\ensuremath{\log{g_{\star}}}}

\newcommand{\mearth}{\ensuremath{M_\earth}}

\newcommand{\rpl}{\ensuremath{R_{p}}}
\newcommand{\mpl}{\ensuremath{M_{p}}}

\newcommand{\rhopl}{\ensuremath{\rho_{p}}}

\newcommand{\arstar}{\ensuremath{a/\rstar}}
\newcommand{\zrstar}{\ensuremath{\zeta/\rstar}}
\newcommand{\rjup}{\ensuremath{R_{\rm J}}}
\newcommand{\mjup}{\ensuremath{M_{\rm J}}}

\newcommand{\refsecl}[1]{\mbox{Section \ref{sec:#1}}}

\newcommand{\reftabl}[1]{Table~\ref{tab:#1}}

\newcommand{\hatcurCCra}{\ensuremath{19^{\mathrm h}39^{\mathrm m}46.08{\mathrm s}}}                                  
\newcommand{\hatcurCCdec}{\ensuremath{-25{\arcdeg}44{\arcmin}53.9{\arcsec}}}                                 
\newcommand{\hatcurCCmag}{14.03}                                       
\newcommand{\hatcurCCtwomass}{2MASS~19394601-2544539}                  
\newcommand{\hatcurCCtassmv}{\ensuremath{14.03\pm0.10}}                
\newcommand{\hatcurCCtwomassJmag}{\ensuremath{13.098\pm0.024}}         
\newcommand{\hatcurCCtwomassHmag}{\ensuremath{12.779\pm0.029}}         
\newcommand{\hatcurCCtwomassKmag}{\ensuremath{12.661\pm0.033}}         
\newcommand{\hatcurLCdip}{\ensuremath{8.3}}                            
\newcommand{\hatcurLCrprstar}{\ensuremath{0.0814\pm0.0038}}            
\newcommand{\hatcurLCbsq}{\ensuremath{0.131_{-0.096}^{+0.192}}}        
\newcommand{\hatcurLCimp}{\ensuremath{0.36_{-0.18}^{+0.21}}}           
\newcommand{\hatcurLCzeta}{\ensuremath{17.42\pm0.32}}                  
\newcommand{\hatcurLCdur}{\ensuremath{0.1263\pm0.0030}}                
\newcommand{\hatcurLCdurhrshort}{\ensuremath{3.031}}                   
\newcommand{\hatcurLCingdur}{\ensuremath{0.0109\pm0.0023}}             
\newcommand{\hatcurLCP}{\ensuremath{3.583893\pm0.000010}}              
\newcommand{\hatcurLCPprec}{\ensuremath{3.5838933}}                    
\newcommand{\hatcurLCPshort}{\ensuremath{3.5839}}                      
\newcommand{\hatcurLCT}{\ensuremath{2456672.1102\pm0.0012}}            
\newcommand{\hatcurSMEiteff}{\ensuremath{5718\pm50}}                   
\newcommand{\hatcurSMEizfeh}{\ensuremath{0.240\pm0.080}}               
\newcommand{\hatcurSMEizfehshort}{\ensuremath{0.24}}                   
\newcommand{\hatcurSMEilogg}{\ensuremath{4.48\pm0.10}}                 
\newcommand{\hatcurSMEivsin}{\ensuremath{1.20\pm0.50}}                 
\newcommand{\hatcurSMEivmac}{\ensuremath{0.0}}                         
\newcommand{\hatcurSMEivmic}{\ensuremath{0.0}}                         
\newcommand{\hatcurSMEiiteff}{\ensuremath{5679\pm50}}                  
\newcommand{\hatcurSMEiizfeh}{\ensuremath{0.210\pm0.080}}              
\newcommand{\hatcurSMEiizfehshort}{\ensuremath{0.21}}                  
\newcommand{\hatcurSMEiilogg}{\ensuremath{4.40\pm0.10}}                
\newcommand{\hatcurSMEiivsin}{\ensuremath{2.00\pm0.50}}                
\newcommand{\hatcurLBii}{\ensuremath{0.2936}}                          
\newcommand{\hatcurLBiii}{\ensuremath{0.3212}}                         
\newcommand{\hatcurLBir}{\ensuremath{0.3906}}                          
\newcommand{\hatcurLBiir}{\ensuremath{0.3083}}                         
\newcommand{\hatcurISOmlong}{\ensuremath{1.056\pm0.037}}               
\newcommand{\hatcurISOrlong}{\ensuremath{1.086_{-0.059}^{+0.149}}}     
\newcommand{\hatcurISOitrho}{\ensuremath{1.20_{-0.29}^{+0.17}}}          
\newcommand{\hatcurISOrholong}{\ensuremath{1.15_{-0.35}^{+0.21}}}      
\newcommand{\hatcurISOlogg}{\ensuremath{4.386\pm0.071}}                
\newcommand{\hatcurISOlum}{\ensuremath{1.11_{-0.13}^{+0.31}}}          
\newcommand{\hatcurISOmv}{\ensuremath{4.73\pm0.20}}                    
\newcommand{\hatcurISOage}{\ensuremath{5.1\pm1.7}}                     
\newcommand{\hatcurISOMK}{\ensuremath{3.14\pm0.19}}                    
\newcommand{\hatcurISOspec}{G}                                         
\newcommand{\hatcurRVK}{\ensuremath{17.7\pm2.5}}                       
\newcommand{\hatcurRVjitter}{\ensuremath{0.00\pm0.99}}                 
\newcommand{\hatcurPPi}{\ensuremath{87.8_{-1.8}^{+1.2}}}               
\newcommand{\hatcurPPlogg}{\ensuremath{2.653_{-0.137}^{+0.099}}}       
\newcommand{\hatcurPPar}{\ensuremath{9.21_{-1.04}^{+0.54}}}            
\newcommand{\hatcurPParel}{\ensuremath{0.04667\pm0.00055}}             
\newcommand{\hatcurPPrho}{\ensuremath{0.259\pm0.091}}                  
\newcommand{\hatcurPPm}{\ensuremath{0.138\pm0.019}}                    
\newcommand{\hatcurPPmlong}{\ensuremath{0.138\pm0.019}}                
\newcommand{\hatcurPPr}{\ensuremath{0.873_{-0.075}^{+0.123}}}          
\newcommand{\hatcurPPrlong}{\ensuremath{0.873_{-0.075}^{+0.123}}}      
\newcommand{\hatcurPPmrcorr}{\ensuremath{-0.08}}                       
\newcommand{\hatcurPPteff}{\ensuremath{1324_{-38}^{+79}}}              
\newcommand{\hatcurPPtheta}{\ensuremath{0.0138\pm0.0026}}              
\newcommand{\hatcurPPfluxavg}{\ensuremath{6.94_{-0.77}^{+1.82}}}       
\newcommand{\hatcurXAv}{\ensuremath{0.0000\pm0.0094}}                  
\newcommand{\hatcurXdistred}{\ensuremath{829_{-48}^{+110}}}            
\newcommand{\hatcurRVeccentwosiglimeccen}{\ensuremath{<0.376}}         
\newcommand{\hatcur}{HATS-8}
\newcommand{\hatcurb}{HATS-8b}
\newcommand{\hatcurRVgammaabs}{\ensuremath{19.958\pm0.041}}                           
\newcommand{\hs}{HATSouth}
\newcommand{\bneighbor}{2MASS19394689-2544386}

\newcommand{\hatcurisoshort}{YY}

\newcommand{\hatcurlumind}{\rhostar}
\newcommand{\hatcurjhkfilset}{ESO}

\newcommand{\hatcurSMEversion}{ii}                                       
\newcommand{\hatcurSMEteff}{\ifthenelse{\equal{\hatcurSMEversion}{i}}{\hatcurSMEiteff}{\hatcurSMEiiteff}}
\newcommand{\hatcurSMEzfeh}{\ifthenelse{\equal{\hatcurSMEversion}{i}}{\hatcurSMEizfeh}{\hatcurSMEiizfeh}}
\newcommand{\hatcurSMEzfehshort}{\ifthenelse{\equal{\hatcurSMEversion}{i}}{\hatcurSMEizfehshort}{\hatcurSMEiizfehshort}}
\newcommand{\hatcurSMElogg}{\ifthenelse{\equal{\hatcurSMEversion}{i}}{\hatcurSMEilogg}{\hatcurSMEiilogg}}
\newcommand{\hatcurSMEvsin}{\ifthenelse{\equal{\hatcurSMEversion}{i}}{\hatcurSMEivsin}{\hatcurSMEiivsin}}
\newcommand{\hatcurSMEvmac}{\ifthenelse{\equal{\hatcurSMEversion}{i}}{\hatcurSMEivmac}{\hatcurSMEiivmac}}
\newcommand{\hatcurSMEvmic}{\ifthenelse{\equal{\hatcurSMEversion}{i}}{\hatcurSMEivmic}{\hatcurSMEiivmic}}

\newboolean{emulateapj}
\setboolean{emulateapj}{true}

\newboolean{rvtablelong}
\setboolean{rvtablelong}{true}

\newboolean{astroph}
\setboolean{astroph}{true}

\submitted{Accepted for publication in The Astronomical Journal; May 29, 2015}
\shortauthors{Bayliss et al.}
\shorttitle{
\hatcur\lowercase{b}
}
\ifthenelse{\boolean{emulateapj}}{
    \newcommand{\titledag}{$\dagger$}
}{
    \newcommand{\titledag}{\dagger}
}

\begin{document}
\title{
\hatcur\lowercase{b}: A Low-Density Transiting Super-Neptune
\altaffilmark{\titledag}
}

\author{ D.~Bayliss\altaffilmark{1,2}, J.~D.~Hartman\altaffilmark{3},
  G.~\'A.~Bakos\altaffilmark{3,$\star$,$\star\star$},
  K.~Penev\altaffilmark{3}, G.~Zhou\altaffilmark{1},
  R.~Brahm\altaffilmark{4,5}, 
  M.~Rabus\altaffilmark{4,6}, A.~Jord\'an\altaffilmark{4,5},
  L.~Mancini\altaffilmark{6},
  M.~de~Val-Borro\altaffilmark{3}, W.~Bhatti\altaffilmark{3}, 
  N.~Espinoza\altaffilmark{4,5}, Z.~Csubry\altaffilmark{3},
  A.~W.~Howard\altaffilmark{7}, B.~J.~Fulton\altaffilmark{7,8},
  L.~A.~Buchhave\altaffilmark{9,10}, T.~Henning\altaffilmark{6},
  B.~Schmidt\altaffilmark{2}, S.~Ciceri\altaffilmark{6},
  R.~W.~Noyes\altaffilmark{9}, H.~Isaacson\altaffilmark{11},
  G.~W.~Marcy\altaffilmark{11},  V.~Suc\altaffilmark{4},
  J.~L\'az\'ar\altaffilmark{12}, I.~Papp\altaffilmark{12},
  P.~S\'ari\altaffilmark{12} } \altaffiltext{1}{Observatoire
  Astronomique de l'Universit\'e de Gen\`eve, 51 ch. des Maillettes,
  1290 Versoix, Switzerland; email: daniel.bayliss@unige.ch}
\altaffiltext{2}{Research School of Astronomy and Astrophysics,
  Australian National University, Canberra, ACT 2611, Australia}
\altaffiltext{3}{Department of Astrophysical Sciences, Princeton
  University, NJ 08544, USA} \altaffiltext{$\star$}{Alfred P.~Sloan
  Research Fellow} \altaffiltext{$\star\star$}{Packard Fellow}
\altaffiltext{4}{Instituto de Astrof\'isica, Facultad de F\'isica,
  Pontificia Universidad Cat\'olica de Chile, Av. Vicu\~na Mackenna
  4860, 7820436 Macul, Santiago, Chile; rbrahm@astro.puc.cl}
\altaffiltext{5}{Millennium Institute of Astrophysics, Av. Vicu\~na
  Mackenna 4860, 7820436 Macul, Santiago, Chile} \altaffiltext{6}{Max
  Planck Institute for Astronomy, Heidelberg, Germany}
\altaffiltext{7}{Institute for Astronomy, University of Hawaii at
  Manoa, Honolulu, HI, USA}
\altaffiltext{8}{NSF Graduate Research Fellow}
\altaffiltext{9}{Harvard-Smithsonian Center for Astrophysics,
  Cambridge, MA 02138, USA}
\altaffiltext{10}{Centre for Star and Planet Formation, Natural History Museum of Denmark,
University of Copenhagen, DK-1350 Copenhagen, Denmark}
\altaffiltext{11}{Department of Astronomy,
  University of California, Berkeley, CA 94720-3411, USA}
\altaffiltext{12}{Hungarian Astronomical
   Association, Budapest, Hungary}
\altaffiltext{$\dagger$}{
The \hs{} network is operated by a collaboration consisting of
Princeton University (PU), the Max Planck Institute f\"ur Astronomie
(MPIA), the Australian National University (ANU), and the Pontificia
Universidad Cat\'olica de Chile (PUC).  The station at Las Campanas
Observatory (LCO) of the Carnegie Institute is operated by PU in
conjunction with PUC, the station at the High Energy Spectroscopic
Survey (H.E.S.S.) site is operated in conjunction with MPIA, and the
station at Siding Spring Observatory (SSO) is operated jointly with
ANU.
This paper includes data gathered with the 6.5\,m Magellan Telescopes
located as Las Campanas Observatory, Chile. Based in part on
observations made with the MPG~2.2\,m Telescope and the ESO~3.6\,m
Telescope at the ESO Observatory in La Silla.  This
paper uses observations obtained with facilities of the Las Cumbres
Observatory Global Telescope.
}


\begin{abstract}

\setcounter{footnote}{10}
\hatcurb{} is a low density transiting super-Neptune discovered as
part of the \hs\ project.  The planet orbits its solar-like G dwarf
host (V=$\hatcurCCtassmv$, \teff=$\hatcurSMEteff$\,K) with a period of
$\hatcurLCPshort$\,d. \hatcurb{} is the third lowest mass transiting
exoplanet to be discovered from a wide-field ground based search, and
with a mass of $\hatcurPPm$\,\mjup{} it is approximately half-way
between the masses of Neptune and Saturn.  However \hatcurb{} has a
radius of \hatcurPPr\,\rjup, resulting in a bulk density of just
$\hatcurPPrho$\,\gcmc.  The metallicity of the host star is
super-Solar (\feh=\hatcurSMEiizfeh), arguing against the idea that low
density exoplanets form from metal-poor environments.  The low density
and large radius of \hatcurb{} results in an atmospheric scale height
of almost 1000\,km, and in addition to this there is an excellent
reference star of near equal magnitude at just 19\arcsec\ separation
on the sky.  These factors make \hatcurb{} an exciting target for
future atmospheric characterization studies, particularly
for long-slit transmission spectroscopy.

\setcounter{footnote}{0}
\end{abstract}



\section{Introduction}
\label{sec:introduction}
While the Solar System planets display a rich diversity of physical
properties, the discovery of exoplanets over the last two decades have
revealed an even wider range of systems.  The mass regimes of
exoplanets is an example of this expanding diversity.  The discovery
of ``super-Earths'' \citep{rivera:2005:gj876}, with masses of
$2<\mearth<10$, have shown us a type of planet unlike anything in our
Solar System.  It now appears that planets in this mass range are
relatively common in the Galaxy \citep{fressin:2013}, although for
many planets discovered by Kepler the masses can only be inferred from
radii in a statistical sense (using mass-radius relationships) due to the
difficulties of measuring radial velocities or transit timing
variations for small planets orbiting faint host stars.

Another mass-class of planets that we only find outside our Solar
System are planets more massive than Neptune at 0.05\mjup, but smaller
than Saturn at 0.3\mjup.  These low mass gas planets, or
``super-Neptunes'' are at the very transition between planets with
H/He dominated compositions and those without H/He dominating the bulk
composition.  In this paper we present the discovery of \hatcurb{}, a
transiting super-Neptune found as part of the \hs\ survey for southern
transiting exoplanets \cite{bakos:2013:hatsouth}.  \hatcurb{} has a
mass lying almost half-way between Neptune and Saturn.  One of the
great advantages of ground-based transit surveys that target bright
stars is that we can measure both the mass and radius of discovered
exoplanets, and therefore we are able to determine the density of the
exoplanet.  This is particularly important at these mass-ranges, as we
cannot infer the radius from the mass alone or vice versa.

In Section~\ref{sec:obs} of this paper we outline the photometric and
spectroscopic observations that led to the detection of \hatcurb.  In
Section~\ref{sec:analysis} we detail the methods we used to determine
the physical parameters of the planet and star as well as ruling out
non-planetary interpretations of the data.  Finally, in Section
\ref{sec:discussion} we put \hatcurb{} into context with other
exoplanets discovered in this mass/density range, and discuss the fact
that this exoplanet orbits a super-solar metallicity star.  We also
discuss the possible follow-up opportunities for this system from the
ground and from space.


\section{Observations}
\label{sec:obs}

\subsection{Photometric detection}
\label{sec:detection}

\begin{figure}[]
  \includegraphics[scale=0.7,angle=0]{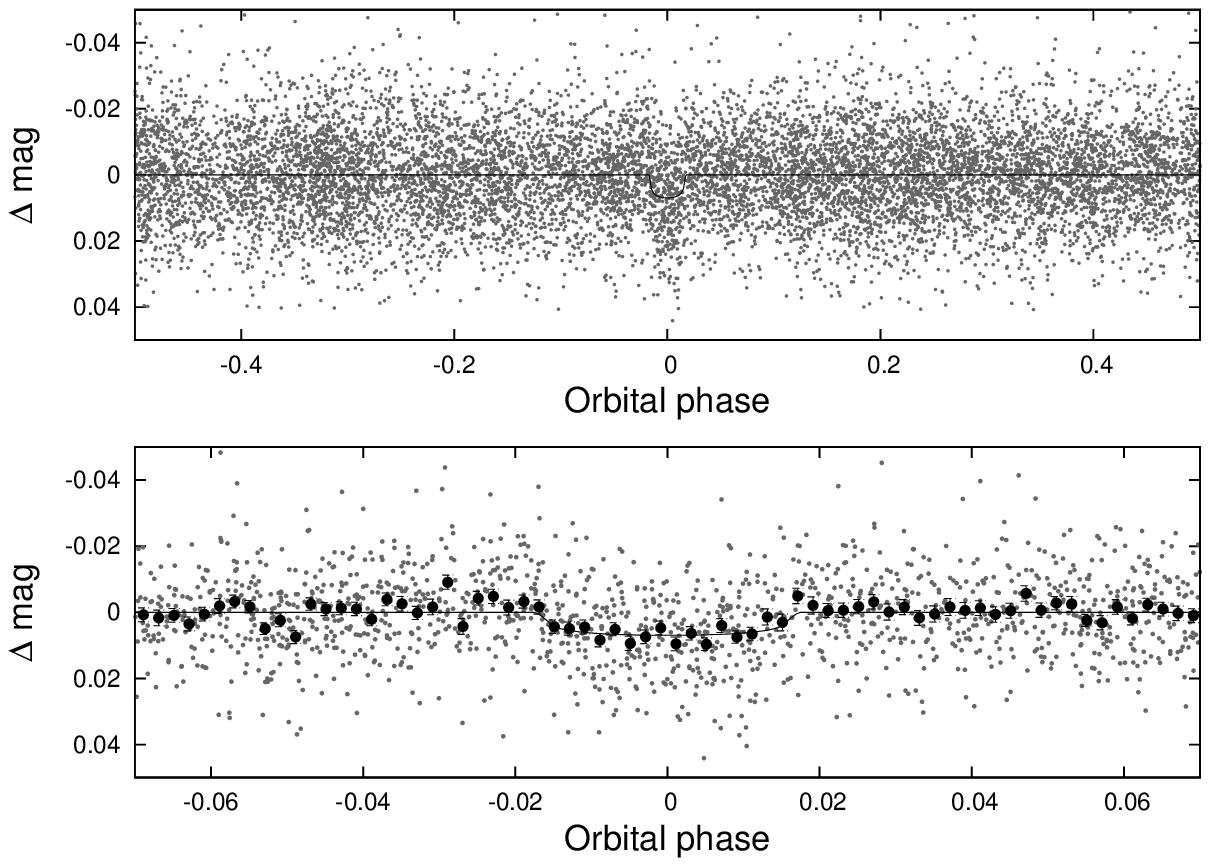}
\caption[]{
        The \hs{} discovery \lc{} for \hatcurb.  Individual points show
        the unbinned instrumental \band{r} \lc{} of \hatcur{} folded
        with the period $P = \hatcurLCPprec$\,days resulting from the
        global fit described in \refsecl{analysis}.  The
        solid line shows the best-fit transit model (see
        \refsecl{analysis}).  The lower panel is a zoom-in on the
        transit feature, which has a depth of \hatcurLCdip~mmag and duration
        of \hatcurLCdurhrshort~hours.  The dark filled points in the lower
        panel show the \lc{} binned in phase using a bin-size of
        0.002.\\
\label{fig:hatsouth}}
\end{figure}

\hatcur{} was intensively monitored as part of the \hs{} survey
\citep{bakos:2013:hatsouth}.  Although some data was acquired as early
as September 2009, the bulk of the observations, over 10,000 images,
were taken between March 2011 and August 2011 (see
Table~\ref{tab:photobs}).  Details of the \hs{} imaging system can be
found in \citet{bakos:2013:hatsouth,penev:2013:hats1}, while here we
provide a summary of the critical features.  \hs{} employs Takahashi
astrographs ($f/2.8$, 18cm apertures) imaged on to Apogee U16M
$4\textrm{K}\times4\textrm{K}$ cameras.  Imaging is performed in the
Sloan \band{r} with exposure times of 240s and a mean cadence of
approximately 300s.  Images are collected from all three sites in the
global network (see Table~\ref{tab:photobs}, which also gives a
breakdown of the number of images taken at each site), and data is
reduced and \lcs{} produced via aperture photometry in the manner
detailed in \cite{penev:2013:hats1}.  \Lcs{} are detrended using
External Parameter Decorrelation \citep[EPD;][]{bakos:2010:hat11} and
the Trend Filtering Algorithm \citep[TFA;][]{kovacs:2005:tfa}).  The
detrended \lc{} for \hatcur{} is set out in Table~\ref{tab:phfu}.  We
search \lcs{} for periodic transit-like events using the Box-fitting
Least Squares algorithm \citep[BLS][]{kovacs:2002:BLS}.  For the case
of \hatcur, we find a transit signal at a period of
P=\hatcurLCPshort~days with a depth of \hatcurLCdip~mmag (see
Figure~\ref{fig:hatsouth}).  No observable out-of-transit variation or
secondary eclipse is apparent.  We therefore began reconnaissance
spectroscopic observations of this candidate as is detailed in
Section~\ref{sec:recon}.

\ifthenelse{\boolean{emulateapj}}{
    \begin{deluxetable*}{llrrrr}
}{
    \begin{deluxetable}{llrrrr}
}
\tablewidth{0pc}
\tabletypesize{\scriptsize}
\tablecaption{
    Summary of photometric observations
    \label{tab:photobs}
}
\tablehead{
    \multicolumn{1}{c}{Facility}          &
    \multicolumn{1}{c}{Date(s)}             &
    \multicolumn{1}{c}{Number of Images}\tablenotemark{a}      &
    \multicolumn{1}{c}{Cadence (s)}\tablenotemark{b}         &
    \multicolumn{1}{c}{Filter}            &
    \multicolumn{1}{c}{Precision (mmag)} \\
    &
    &
    &
    &
    &
}
\startdata
~~~~HS-1/G579 & 2009 Sep--2011 Aug & 4484 & 301 & \band{r} & 13.9 \\
~~~~HS-3/G579 & 2010 Mar--2011 Aug & 2607 & 303 & \band{r} & 13.3 \\
~~~~HS-5/G579 & 2010 Sep--2011 Aug & 3303 & 303 & \band{r} & 12.2 \\
~~~~Swope/SiTe3 & 2013 May 29 & 107 & 213 & \band{i} & 2.3 \\
~~~~Swope/E2V & 2014 Jul 01 & 116 & 189 & \band{i} & 1.8 \\
~~~~LCOGT~1m/Sinistro & 2014 Jul 09 & 50 & 347 & \band{i} & 3.5 \\
[-1.5ex]
\enddata
\tablenotetext{a}{
  Excludes images which were rejected as significant outliers in the
  fitting procedure.
}
\tablenotetext{b}{
  The mode time difference rounded to the nearest second between
  consecutive points in each \lc.  Due to visibility, weather,
  pauses for focusing, etc., none of the \lcs{} have perfectly
  uniform time sampling.
}
\ifthenelse{\boolean{emulateapj}}{
    \end{deluxetable*}
}{
    \end{deluxetable}
}

\subsection{Reconnaissance Spectroscopy}
\label{sec:recon}
Although a periodic dip in the \lc{} of a star may be due to a
transiting exoplanet, there are many other astrophysical sources of
such a signal, which can be termed ``false positives''.  Blended
eclipsing binaries and grazing eclipsing binaries are common causes of
false positives.  Other candidates can be ``false alarms''; that is
the detected event is non-astrophysical in nature.  These can be
caused by artifacts on the CCD or period systematics introduced by the
telescope or mount systems.  The combination of all of these false
candidates, coupled with a desire to find shallow transits
(i.e. smaller radius planets), means that the majority of candidates
produced by ground-based transit surveys such as the \hs{} survey are
not genuine transiting exoplanets.  It is therefore crucial to
undertake an efficient vetting program of transiting exoplanet candidates by
way of reconnaissance spectroscopy before attempting the resource
intensive search for subtle radial velocity variations consistent
with an orbiting exoplanet.  Such reconnaissance spectroscopy will
become even more critical as large surveys such as TESS
\citep{ricker:2014:TESS} and PLATO \citep{rauer:2014:PLATO} produce
enormous numbers of candidates requiring such vetting.

The candidate \hatcurb{} is a good
example of low a signal-to-noise detection, with a transit depth in
the discovery \lc{} of just 8.3\,mmag, compared to an RMS precision of
the \lc{} per observation of 12 to 14\,mmag level (see
Table~\ref{tab:photobs}).  

In the case of \hatcur{}, follow-up spectroscopy was performed in
October 2012 using the echelle spectrograph on the 2.5\,m Du Pont
telescope at Las Campanas Observatory in Chile.  We obtained two
spectra using the $1\arcsec\times4\arcsec$ slit (R$\sim$40000) on the
nights of October 25 and October 26 2012, each with an exposure time
of 1800\,s.  From these observations we calculated that \hatcur{} was
a G-dwarf with a low projected rotational velocity ($\vsini$) and
there was no sign of a secondary spectrum that could be indicative of
a binary system.  A further observation of \hatcur{} was obtained with
the FEROS spectrograph \citep{kaufer:1998} on the MPG 2.2\,m telescope
at the ESO Observatory in La Silla, Chile.  This instrument, with
slightly higher spectral resolution, confirmed the Du Pont finding
that \hatcur{} is a G-dwarf with low $\vsini$ and no evidence of a
composite spectrum.  The three observations also showed no radial
velocity variation to within the uncertainty of the measurements,
indicating the system could not be a eclipsing binary system with a
large radial velocity amplitude.  Details of the three reconnaissance
spectra taken for \hatcur{} are set out in
Table~\ref{tab:specobssummary}.  Based on the reconnaissance
spectroscopy, \hatcur{} was deemed to be a sufficiently strong
candidate to warrant undertaking precise photometric and radial
velocity measurements as detailed in Sections~\ref{sec:pphfu} and
\ref{sec:hispec} respectively.

\subsection{Precise Photometric Follow-up}
\label{sec:pphfu}
\begin{figure*}[!ht]
\plotone{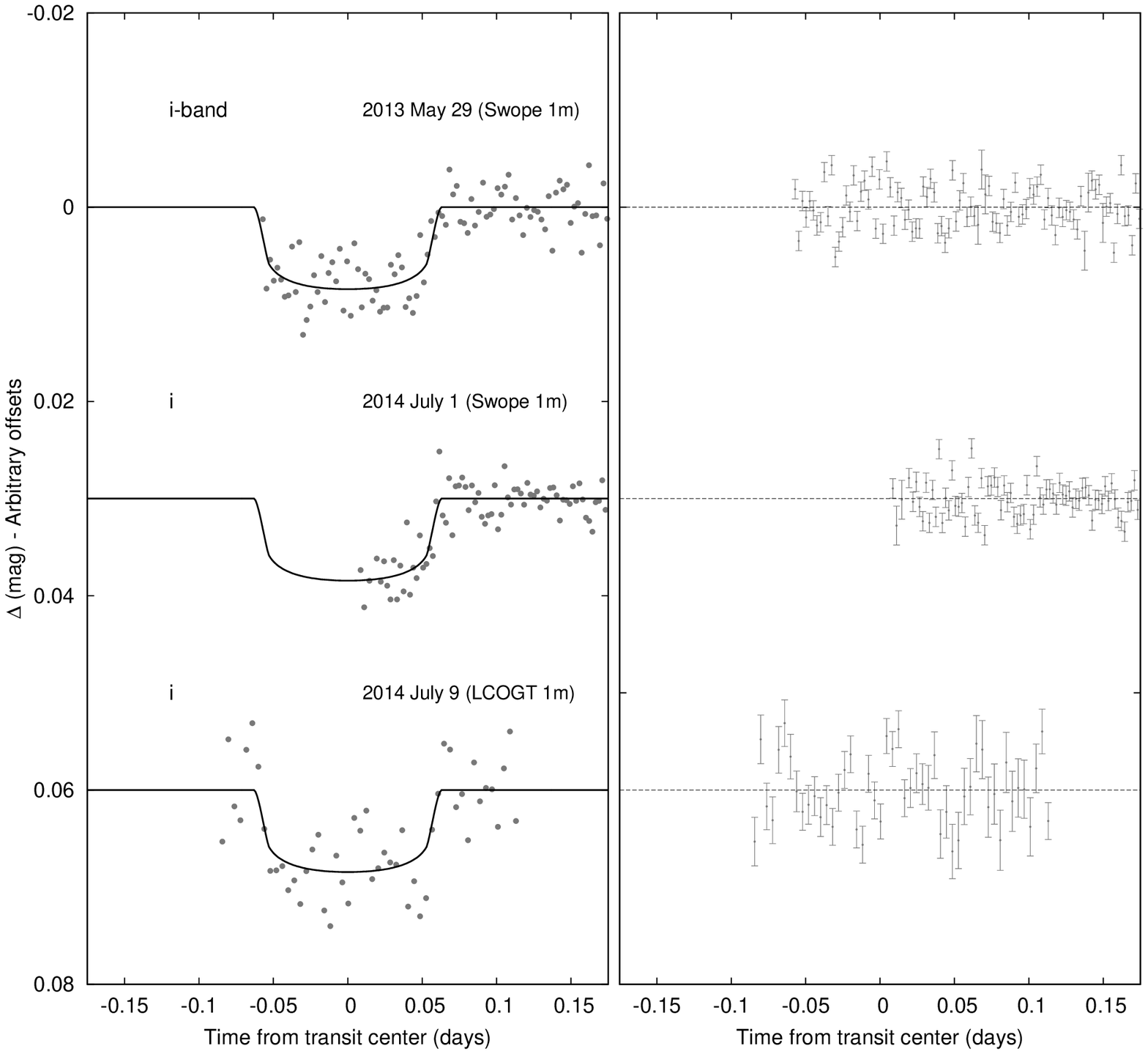}
\caption{
  Left: Follow-up \band{i} transit \lcs{} of \hatcur{} from the
  Swope~1\,m and LCOGT~1\,m telescopes in Chile.  The dates and
  instruments used for each event are indicated.  The \lcs{} have been
  detrended using the EPD process.  \Lcs{} after the first are shifted
  for clarity.  Our best fit is shown by the solid lines.  Right:
  Residuals from the fits in the same order as the curves at left.\\
\label{fig:lc}} \end{figure*}

Precise photometric observations of a transit of \hatcurb{} were
carried out using the SiTe3 imaging camera on the SWOPE 1m telescope
at Las Campanas Observatory (LCO) in Chile on 2013 May 29.  Defocused
imaging over the 14.8$\arcmin\times$22.8$\arcmin$ field was performed
in \band{i} with exposure times of 120s.  The images were reduced and
aperture photometry extracted in the standard manner as set out in
\citet{penev:2013:hats1}.  Figure~\ref{fig:lc} shows the resulting
\lc, which covers all but the ingress of the transit on that night,
and allows for a very precise determination of the period and phase of
\hatcurb{}.  Using this precise ephemeris, a second photometric
observations of transits of \hatcurb{} was obtained with the 1m SWOPE
telescope on the night of 2014 July 1.  Again the monitoring was
carried out in \band{i}, this time using the E2V imaging camera with
exposure times of 150s.  The egress of the transit of \hatcurb{} was
detected with these observations, as detailed in Figure~\ref{fig:lc}.
Finally, nine days later we observed another transit of \hatcurb{},
this time using the LCOGT 1m telescope network
\citep{brown:2013:lcogt}.  This observation was carried out with the
LCOGT\,1m telescope at CTIO in Chile using the Sinistro imaging camera
in \band{i} with a 300\,s exposure time and the telescope slightly
defocused.  The observation covered a full transit.  The transits are
shown in Figure~\ref{fig:lc}.

All three transit events monitored with higher precision photometry
confirmed the transit signal detected in the \hs{} discovery \lcs.
The depth and shape of the transit are consistent with that expected
from a transiting exoplanet, and we use these data in combination with
the discovery photometry to determine the global parameters of the
\hatcurb{} system in Section~\ref{sec:analysis}.

\ifthenelse{\boolean{emulateapj}}{
        \begin{deluxetable*}{lrrrrr} }{
        \begin{deluxetable}{lrrrrr} 
    }
        \tablewidth{0pc}
        \tablecaption{Differential photometry of
        \hatcur\label{tab:phfu}} \tablehead{ \colhead{BJD} &
        \colhead{Mag\tablenotemark{a}} &
        \colhead{\ensuremath{\sigma_{\rm Mag}}} &
        \colhead{Mag(orig)\tablenotemark{b}} & \colhead{Filter} &
        \colhead{Instrument} \\ \colhead{\hbox{~~~~(2\,400\,000$+$)~~~~}}
        & \colhead{} & \colhead{} & \colhead{} & \colhead{} &
        \colhead{} } \startdata
        $ 55774.34619 $ & $   0.00664 $ & $   0.00797 $ & $ \cdots $ & $ r$ &         HS\\
$ 55699.08460 $ & $  -0.02530 $ & $   0.01660 $ & $ \cdots $ & $ r$ &         HS\\
$ 55709.83639 $ & $  -0.00581 $ & $   0.02028 $ & $ \cdots $ & $ r$ &         HS\\
$ 55770.76289 $ & $  -0.02414 $ & $   0.00927 $ & $ \cdots $ & $ r$ &         HS\\
$ 55763.59560 $ & $  -0.00056 $ & $   0.00983 $ & $ \cdots $ & $ r$ &         HS\\
$ 55788.68300 $ & $  -0.00119 $ & $   0.01462 $ & $ \cdots $ & $ r$ &         HS\\
$ 55763.59611 $ & $  -0.01683 $ & $   0.01040 $ & $ \cdots $ & $ r$ &         HS\\
$ 55691.91833 $ & $  -0.04584 $ & $   0.01747 $ & $ \cdots $ & $ r$ &         HS\\
$ 55663.24726 $ & $  -0.01226 $ & $   0.00956 $ & $ \cdots $ & $ r$ &         HS\\
$ 55727.75748 $ & $   0.02984 $ & $   0.02475 $ & $ \cdots $ & $ r$ &         HS\\
        [-1.5ex]
\enddata \tablenotetext{a}{
     The out-of-transit level has been subtracted. For the \hs{}
     \lc{} (rows with ``HS'' in the Instrument column), these
     magnitudes have been detrended using the EPD and TFA procedures
     prior to fitting a transit model to the \lc. Primarily as
     a result of this detrending, but also due to blending from
     neighbors, the apparent \hs{} transit depth is somewhat
     shallower than that of the true depth in the Sloan~$r$ filter
     (the apparent depth is 85\% that of the true depth). For the
     follow-up \lcs{} (rows with an Instrument other than
     ``HS'') these magnitudes have been detrended with the EPD
     procedure, carried out simultaneously with the transit fit (the
     transit shape is preserved in this process).
}
\tablenotetext{b}{
        Raw magnitude values without application of the EPD
        procedure.  This is only reported for the follow-up \lcs.
}
\tablecomments{
        This table is available in a machine-readable form in the
        online journal.  A portion is shown here for guidance
        regarding its form and content. The data are also available on
        the \hs{} website at \url{http://www.hatsouth.org}.
} \ifthenelse{\boolean{emulateapj}}{ \end{deluxetable*} }{ \end{deluxetable} }

\subsection{High Precision Spectroscopy}
\label{sec:hispec}

Radial velocity measurements play a critical role in the confirmation
and characterization of transiting planets, and are needed to determine the
mass, and hence the bulk density, of the exoplanet.  \hatcur{} is a
difficult prospect for radial velocity measurements, as it is
relatively faint (V=\hatcurCCmag).  For this reason we needed to use
Keck/HIRES in order to have the necessary precision and
signal-to-noise to measure the radial velocity variations.

\begin{figure} [ht]
\plotone{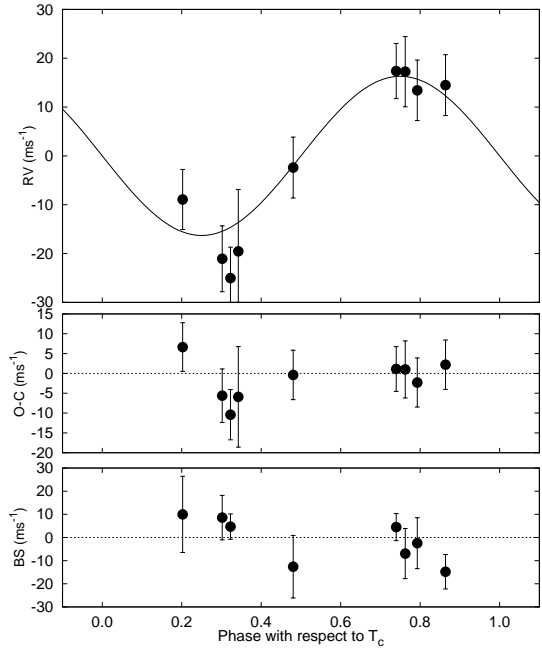}
\caption{
    {\em Top panel:} High-precision radial velocity measurements for
    \hbox{\hatcur{}} from Keck/HIRES,
    together with our best-fit circular orbit model.  Zero phase
    corresponds to the time of mid-transit.  The center-of-mass
    velocity has been subtracted.  {\em Second panel:} Velocity
    $O\!-\!C$ residuals from the best-fit model.  The error bars for
    each instrument include the jitter which is varied in the fit.
    {\em Third panel:} Bisector spans (BS), with the mean value
    subtracted.  Note the different vertical scales of the panels.\\
\label{fig:rvbis}}
\end{figure}

We observed \hatcur{} with HIRES \citep{vogt:1994:hires} on Keck-I in
Hawaii between June and September 2014.  We used HIRES in its standard
configuration for precise radial velocity measurements: a slit width
of 0.86\arcsec, $\lambda / \Delta \lambda \approx 55,000$, and
wavelength coverage of ∼3800-8000\,\AA.  Exposure times were typically
1500\,s and achieved a signal-to-noise ratio of 40 per pixel in the
continuum near 5500\,\AA\ in the reduced HIRES spectra.  An iodine gas
absorption cell in the optical path is used to superimpose iodine
absorption lines on the stellar spectrum and provide an accurate
wavelength calibration \cite[see][]{marcy:1992}.  A synthetic
iodine-free template spectrum is created by interpolating the
\citet{coelho:2014} grid of stellar models for the values of \teff,
\logg, \feh{} derived in Section~\ref{sec:host}.  We found that using
the synthetic template produced smaller residuals and measurement
uncertainties when compared to using an observed iodine-free spectrum
with moderate S/N as the template. In the case of the synthetic
template, no deconvolution is needed to remove the effects of an
asymmetric instrumental PSF, and it is completely noise-free. See
\citet{fulton:2015} for a more detailed description and performance
tests of the synthetic template technique.  We derived relative radial
velocity measurement for \hatcur{} using the method described in
Butler et al. (1996), which accounts for variations of the
spectrograph instrumental line profile.  The radial velocity
measurements are listed in Table~\ref{tab:rvs} and plotted in
Figure~\ref{fig:rvbis}, along with the estimated uncertainties in
these measurements and the bisector spans of the average spectral
line.

We computed spectral line bisector spans from the blue orders of the
Keck/HIRES observations following \citet{torres:2007:hat3} and
corrected these for contamination from scattered moonlight following
\citet{hartman:2011:hat1819}.  These data are also represented
graphically as a function of orbital phase in Figure~\ref{fig:rvbis},
along with the best fit circular orbit.  No systematic variation is
seen in the bisector span measurements, which would have been a
signature of a blended system.  The overall scatter of the bisector
spans is only 9.5\,\ms. This is good evidence that \hatcur{} is not
a blended stellar eclipsing binary system. A more thorough rejection
of blends is discussed is Section~\ref{sec:blends}.

We measured the emission in the cores of the Ca~$II$~H\&K lines and
found a \logrhk\ value of -5.179.  This value suggests that this star
has little chromospheric activity. The residuals of the radial
velocity fit show an RMS of 5.1\,\ms, which is consistent with the
formal uncertainties (see Figure~\ref{fig:rvbis}) and implies a very
faint, chromospherically quiet star.

\ifthenelse{\boolean{emulateapj}}{
    \begin{deluxetable*}{llrrrrr}
}{
    \begin{deluxetable}{llrrr}
}
\tablewidth{0pc}
\tabletypesize{\scriptsize}
\tablecaption{
    Summary of spectroscopic observations\label{tab:specobssummary}
}
\tablehead{
    \multicolumn{1}{c}{Telescope/Instrument} &
    \multicolumn{1}{c}{Date Range}          &
    \multicolumn{1}{c}{Number of Observations} &
    \multicolumn{1}{c}{Resolution}          &
    \multicolumn{1}{c}{Observing Mode}          \\
}
\startdata
du~Pont~2.5\,m/Echelle & 2012 Oct 25--26 & 2 & 40000 & RECON \\
MPG~2.2\,m/FEROS & 2013 July 18 & 1 & 48000 & RECON \\
Keck-I~10\,m/HIRES & 2014 June 20 & 1 & 55000 & I2-free template \\
Keck-I~10\,m/HIRES & 2014 June 17 -- 2014 Sep 10 & 9 & 55000 & I2/HPRV \\
[-1.5ex]
\enddata 
\ifthenelse{\boolean{emulateapj}}{
    \end{deluxetable*}
}{
    \end{deluxetable}
}

\ifthenelse{\boolean{emulateapj}}{
    \begin{deluxetable*}{lrrrrrr}
}{
    \begin{deluxetable}{lrrrrrr}
}
\tablewidth{0pc}
\tablecaption{
    Relative radial velocities and bisector span measurements of
    \hatcur{}.
    \label{tab:rvs}
}
\tablehead{
    \colhead{BJD} & 
    \colhead{RV\tablenotemark{a}} & 
    \colhead{\ensuremath{\sigma_{\rm RV}}\tablenotemark{b}} & 
    \colhead{BS} & 
    \colhead{\ensuremath{\sigma_{\rm BS}}} & 
        \colhead{Phase} &
        \colhead{Instrument}\\
    \colhead{\hbox{(2\,456\,000$+$)}} & 
    \colhead{(\ms)} & 
    \colhead{(\ms)} &
    \colhead{(\ms)} &
    \colhead{} &
        \colhead{} &
        \colhead{}
}
\startdata
$ 826.94201 $ & $    -8.93 $ & $     6.15 $ & $   10.0 $ & $   16.4 $ & $   0.202 $ & Keck \\
$ 827.93901 $ & $    -2.39 $ & $     6.24 $ & $  -12.6 $ & $   13.5 $ & $   0.480 $ & Keck \\
$ 828.95001 $ & $    17.25 $ & $     7.19 $ & $   -7.0 $ & $   10.8 $ & $   0.762 $ & Keck \\
$ 830.95801 $ & $   -25.03 $ & $     6.34 $ & $    4.7 $ & $    5.5 $ & $   0.323 $ & Keck \\
$ 874.03501 $ & $   -19.55 $ & $    12.68 $ & \nodata      & \nodata      & $   0.342 $ & Keck \\
$ 882.81701 $ & $    13.43 $ & $     6.20 $ & $   -2.5 $ & $   11.0 $ & $   0.793 $ & Keck \\
$ 889.79501 $ & $    17.38 $ & $     5.65 $ & $    4.5 $ & $    5.9 $ & $   0.740 $ & Keck \\
$ 891.80901 $ & $   -21.07 $ & $     6.75 $ & $    8.6 $ & $    9.6 $ & $   0.302 $ & Keck \\
$ 911.74301 $ & $    14.49 $ & $     6.23 $ & $  -14.8 $ & $    7.4 $ & $   0.864 $ & Keck \\
    [-1.5ex]
\enddata
\tablenotetext{a}{
        The zero-point of these velocities is arbitrary. An overall
        offset $\gamma_{\rm rel}$ fitted separately to the Keck/HIRES
        velocities in \refsecl{analysis} has been subtracted.
}
\tablenotetext{b}{
        Internal errors excluding the component of
        astrophysical/instrumental jitter considered in
        \refsecl{analysis}.
}
\ifthenelse{\boolean{emulateapj}}{
    \end{deluxetable*}
}{
    \end{deluxetable}
}

\section{Analysis}
\label{sec:analysis}
\subsection{Stellar Properties of \hatcur}
\label{sec:host}
The analysis of the reconnaissance spectra, described in
Section~\ref{sec:recon}, indicated that the star \hatcur{} is a slowly
rotating G-dwarf.  We derive more precise stellar parameters by
analyzing the iodine-free template spectra taken with Keck/HIRES on
2014 June 20 using the Stellar Parameter Classification (SPC) method
as described in the supplementary information to
\citet{buchhave:2012}.  This method involves cross-correlating the
observed spectrum between 5050 and 5360 \AA{} with a grid of synthetic
templates covering a wide range of \teff, \vsini, \logg, and \feh.

\begin{figure}[h]
\plotone{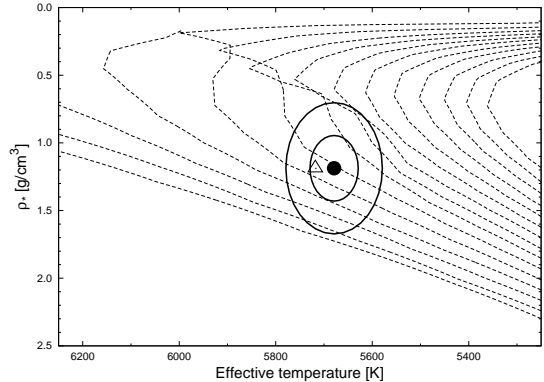}
\caption[]{
    Comparison between the measured values of \teffstar\ and
    \rhostar\ (from SPC applied to the HIRES spectra, and from our
    modeling of the \lcs{} and radial velocity data, respectively), and the
    Y$^{2}$ model isochrones from \citet{yi:2001}. The best-fit values
    (dark filled circle), and approximate 1$\sigma$ and 2$\sigma$
    confidence ellipsoids are shown. The values from our initial SPC
    iteration are shown with the open triangle. The Y$^{2}$ isochrones
    are shown for ages of 0.2\,Gyr, and 1.0 to 14.0\,Gyr in 1\,Gyr
    increments.\\
\label{fig:iso}}
\end{figure}

Using the method described in \citet{sozzetti:2007} and applied in
previous \hs{} discoveries \cite[e.g.][]{penev:2013:hats1}, we
initially derive the mean stellar density of \hatcur{} via \lc{}
fitting, which we determine to be \hatcurISOitrho\,\gcmc. To determine
other stellar properties, we combine this mean stellar density with
the \teffstar{} from SPC analysis, and the Yonsei-Yale
\citep[Y2;][]{yi:2001} stellar evolution models.  This yields a
\loggstar=4.40$\pm$0.06, which we then fix in a second iteration of
SPC to derive the final parameters for \hatcur{}.  The final
parameters are listed in Table~\ref{tab:stellar}, and reveal \hatcur{}
to be \hatcurISOspec\ dwarf host with a \teff=\hatcurSMEteff\,K, a
\logg=\hatcurISOlogg, and a \vsini=\hatcurSMEvsin\,\kms.  We note that
this stellar surface gravity (\logg=\hatcurISOlogg) is consistent with
the value measured from spectra alone (\logg=\hatcurSMEilogg).  We estimate
the age of this system using using the Y2 model isochrones to be
\hatcurISOage~Gyr; see Figure~\ref{fig:iso}.  All of these stellar
parameters make \hatcur{} very similar to our own Sun.  The primary
difference is that the metallicity of \hatcur{} is super-Solar at
\feh=+\hatcurSMEiizfeh.

\subsection{Excluding Blending Scenarios}
\label{sec:blends}
One of the most difficult false positive scenarios to rule out is a
background blended eclipsing binary, which is not obvious in either
photometric or spectroscopic follow-up.  Here we follow the method set
out in \citet{hartman:2011:hat3233} in order to determine the if the
data can be explained by a blended stellar system rather than a
transiting exoplanet.

We find that a model consisting of a single star with a transiting
planet provides a lower $\chi^2$ fit to the photometric data than any
of the blended eclipsing binary models that we tested. The best-fit
blended eclipsing binary model can be rejected in favor of the
star+planet model, based solely on the photometry, with greater than
$2\sigma$ confidence. We also simulated cross-correlation functions
(CCFs), radial velocities and bisector span measurements for each
blended eclipsing binary model considered, and find that all of the
models that come close to fitting the photometric data would have
easily been detected as composite stellar systems based on the
spectroscopic observations (they have clearly double-peaked CCFs
and/or radial velocities and bisector spans that vary by more than
1\,\kms). We therefore conclude that \hatcur{} is a transiting planet
system.

We also considered the possibility that \hatcur{} is a transiting
planet system with an unresolved stellar companion next to the host star. While we cannot
rule out very low mass stellar companions, we find that a companion
with $M \ga 0.65$\,\msun\ can be excluded with greater than $3\sigma$
confidence. If \hatcur{} has a $0.65$\,\msun\ companion, the radius of
\hatcurb{} would be $\sim 6$\% larger than what we measure assuming no
stellar companion; the mass would also be slightly larger.

\subsection{Global Modeling of Data}
\label{sec:global}
We derive the parameters for the \hatcurb{} system using a joint,
Markov-Chain Monte Carlo method detailed in \cite{bakos:2010:hat11},
and with modifications set out in \cite{hartman:2012:hat39hat41}.
This method models the best fit parameters in a global way using the
discovery photometry, follow-up photometry, and precise radial
velocity measurements.  The resulting planetary parameters are set out
in detail in Table~\ref{tab:planetparam}.  We model the system for
both a circular planet orbit and for an orbit where eccentricity is
allowed to vary.  We calculate the Bayesian evidence for both these
scenarios, and find that the circular orbit is preferred.  However we
also list the $95\%$ confidence upper limit, which is
$e$\hatcurRVeccentwosiglimeccen{}.

Our global modeling reveals that \hatcurb{} is a \hatcurPPm\,\mjup{}
exoplanet, which is just 2.5 times the mass of Neptune or one-third
the mass of Saturn.  The radius is \hatcurPPr\,\rjup, which results in
a bulk density of just \rhopl=\hatcurPPrho\,\gcmc.  Its orbital
distance of \hatcurPParel\,AU implies that \hatcurb{} would have a
temperature of \hatcurPPteff{}\,K (assuming zero albedo and a complete
redistribution of heat).

\ifthenelse{\boolean{emulateapj}}{
  \begin{deluxetable*}{lcr}
}{
  \begin{deluxetable}{lcr}
}
\tablewidth{0pc}
\tabletypesize{\scriptsize}
\tablecaption{
    Stellar Parameters for \hatcur{} 
    \label{tab:stellar}
}
\tablehead{
    \multicolumn{1}{c}{~~~~~~~~Parameter~~~~~~~~} &
    \multicolumn{1}{c}{Value}                     &
    \multicolumn{1}{c}{Source}    
}
\startdata
\noalign{\vskip -3pt}
\sidehead{Identifying Information}
~~~~R.A. (h:m:s)                      &  \hatcurCCra{} & 2MASS\\
~~~~Dec. (d:m:s)                      &  \hatcurCCdec{} & 2MASS\\
~~~~2MASS ID                          &  \hatcurCCtwomass{} & 2MASS\\
\sidehead{Spectroscopic properties}
~~~~$\teffstar$ (K)\dotfill         &  \hatcurSMEteff{} & SPC \tablenotemark{a}\\
~~~~$\feh$\dotfill                  &  \hatcurSMEzfeh{} & SPC                 \\
~~~~$\vsini$ (\kms)\dotfill         &  \hatcurSMEvsin{} & SPC                 \\
~~~~$\gamma_{\rm RV}$ (\kms)\dotfill&  \hatcurRVgammaabs{} & FEROS                  \\
~~~~$\logrhk$ \dotfill          &  $-5.18\pm 0.1$ & KECK                    \\
\sidehead{Photometric properties}
~~~~$V$ (mag)\dotfill               &  \hatcurCCtassmv{} & NOMAD               \\
~~~~$J$ (mag)\dotfill               &  \hatcurCCtwomassJmag{} & 2MASS           \\
~~~~$H$ (mag)\dotfill               &  \hatcurCCtwomassHmag{} & 2MASS           \\
~~~~$K_s$ (mag)\dotfill             &  \hatcurCCtwomassKmag{} & 2MASS           \\
\sidehead{Derived properties}
~~~~$\mstar$ ($\msun$)\dotfill      &  \hatcurISOmlong{} & Isochrones+\hatcurlumind{}+SPC \tablenotemark{b}\\
~~~~$\rstar$ ($\rsun$)\dotfill      &  \hatcurISOrlong{} & Isochrones+\hatcurlumind{}+SPC         \\
~~~~$\loggstar$ (cgs)\dotfill       &  \hatcurISOlogg{} & Isochrones+\hatcurlumind{}+SPC         \\
~~~~$\rhostar$ (cgs)\dotfill        &  \hatcurISOrholong{} & Isochrones+\hatcurlumind{}+SPC \tablenotemark{c}\ \\
~~~~$\lstar$ ($\lsun$)\dotfill      &  \hatcurISOlum{} & Isochrones+\hatcurlumind{}+SPC         \\
~~~~$M_V$ (mag)\dotfill             &  \hatcurISOmv{} & Isochrones+\hatcurlumind{}+SPC         \\
~~~~$M_K$ (mag,\hatcurjhkfilset{})&  \hatcurISOMK{} & Isochrones+\hatcurlumind{}+SPC         \\
~~~~Age (Gyr)\dotfill               &  \hatcurISOage{} & Isochrones+\hatcurlumind{}+SPC         \\
~~~~$A_{V}$ (mag) \tablenotemark{d}\dotfill           &  \hatcurXAv{} & Isochrones+\hatcurlumind{}+SPC\\
~~~~Distance (pc)\dotfill           &  \hatcurXdistred{} & Isochrones+\hatcurlumind{}+SPC\\
\enddata
\tablenotetext{a}{
    SPC = ``Stellar Parameter Classification'' method based on
    cross-correlating high-resolution spectra against synthetic
    templates \citep{buchhave:2012}. These parameters rely primarily
    on SPC, but have a small dependence also on the iterative analysis
    incorporating the isochrone search and global modeling of the
    data, as described in the text.  } 
\tablenotetext{b}{
    Isochrones+\hatcurlumind{}+SPC = Based on the Y$^{2}$ isochrones
    \citep{yi:2001},
    the stellar density used as a luminosity indicator, and the SPC
    results.
} 
\tablenotetext{c}{
  In the case of $\rhostar$ the parameter is primarily determined from
  the global fit to the \lcs\ and radial velocity data. The value shown here
  also has a slight dependence on the stellar models and SPC
  parameters due to restricting the posterior distribution to
  combinations of $\rhostar$+$\teffstar$+$\feh$ that match to a
  \hatcurisoshort{} stellar model.
} 
\tablenotetext{d}{ Total \band{V} extinction to the star determined
  by comparing the catalog broad-band photometry listed in the table
  to the expected magnitudes from the
  Isochrones+\hatcurlumind{}+SPC model for the star. We use the
  \citet{cardelli:1989} extinction law.  }
\ifthenelse{\boolean{emulateapj}}{
  \end{deluxetable*}
}{
  \end{deluxetable}
}
\ifthenelse{\boolean{emulateapj}}{
  \begin{deluxetable*}{lc}
}{
  \begin{deluxetable}{lc}
}
\tabletypesize{\scriptsize}
\tablecaption{Parameters for the transiting planet \hatcurb{}.\label{tab:planetparam}}
\tablehead{
    \multicolumn{1}{c}{~~~~~~~~Parameter~~~~~~~~} &
    \multicolumn{1}{c}{Value \tablenotemark{a}}                     
}
\startdata
\noalign{\vskip -3pt}
\sidehead{\Lc{} parameters}
~~~$P$ (days)             \dotfill    & $\hatcurLCP{}$              \\
~~~$T_c$ (${\rm BJD}$)    
      \tablenotemark{b}   \dotfill    & $\hatcurLCT{}$              \\
~~~$T_{14}$ (days)
      \tablenotemark{b}   \dotfill    & $\hatcurLCdur{}$            \\
~~~$T_{12} = T_{34}$ (days)
      \tablenotemark{b}   \dotfill    & $\hatcurLCingdur{}$         \\
~~~$\arstar$              \dotfill    & $\hatcurPPar{}$             \\
~~~$\zrstar$ \tablenotemark{c}              \dotfill    & $\hatcurLCzeta{}$\phn       \\
~~~$\rpl/\rstar$          \dotfill    & $\hatcurLCrprstar{}$        \\
~~~$b^2$                  \dotfill    & $\hatcurLCbsq{}$            \\
~~~$b \equiv a \cos i/\rstar$
                          \dotfill    & $\hatcurLCimp{}$           \\
~~~$i$ (deg)              \dotfill    & $\hatcurPPi{}$\phn         \\

\sidehead{Limb-darkening coefficients \tablenotemark{d}}
~~~$c_1,i$ (linear term)  \dotfill    & $\hatcurLBii{}$            \\
~~~$c_2,i$ (quadratic term) \dotfill  & $\hatcurLBiii{}$           \\
~~~$c_1,r$               \dotfill    & $\hatcurLBir{}$             \\
~~~$c_2,r$               \dotfill    & $\hatcurLBiir{}$            \\

\sidehead{RV parameters}
~~~$K$ (\ms)              \dotfill    & $\hatcurRVK{}$\phn\phn      \\
~~~$e$ \tablenotemark{e}  \dotfill    & $\hatcurRVeccentwosiglimeccen{}$ \\
~~~RV jitter (\ms) \tablenotemark{f}        \dotfill    & \hatcurRVjitter{}           \\

\sidehead{Planetary parameters}
~~~$\mpl$ ($\mjup$)       \dotfill    & $\hatcurPPmlong{}$          \\
~~~$\rpl$ ($\rjup$)       \dotfill    & $\hatcurPPrlong{}$          \\
~~~$C(\mpl,\rpl)$
    \tablenotemark{g}     \dotfill    & $\hatcurPPmrcorr{}$         \\
~~~$\rhopl$ (\gcmc)       \dotfill    & $\hatcurPPrho{}$            \\
~~~$\log g_p$ (cgs)       \dotfill    & $\hatcurPPlogg{}$           \\
~~~$a$ (AU)               \dotfill    & $\hatcurPParel{}$          \\
~~~$T_{\rm eq}$ (K) \tablenotemark{h}        \dotfill   & $\hatcurPPteff{}$           \\
~~~$\Theta$ \tablenotemark{i} \dotfill & $\hatcurPPtheta{}$         \\
~~~$\langle F \rangle$ ($10^{9}$\ergscmsq) \tablenotemark{i}
                          \dotfill    & $\hatcurPPfluxavg{}$       \\ [-1.5ex]
\enddata
\tablenotetext{a}{
    The adopted parameters assume a circular orbit. Based on the
    Bayesian evidence ratio we find that this model is strongly
    preferred over a model in which the eccentricity is allowed to
    vary in the fit. For each parameter we give the median value and
    68.3\% (1$\sigma$) confidence intervals from the posterior
    distribution.
}
\tablenotetext{b}{
    Reported times are in Barycentric Julian Date calculated directly
    from UTC, {\em without} correction for leap seconds.
    \ensuremath{T_c}: Reference epoch of mid transit that
    minimizes the correlation with the orbital period.
    \ensuremath{T_{14}}: total transit duration, time
    between first to last contact;
    \ensuremath{T_{12}=T_{34}}: ingress/egress time, time between first
    and second, or third and fourth contact.
}
\tablenotetext{c}{
    Reciprocal of the half duration of the transit used as a jump
    parameter in our MCMC analysis in place of $\arstar$. It is
    related to $\arstar$ by the expression $\zrstar = \arstar
    (2\pi(1+e\sin \omega))/(P \sqrt{1 - b^{2}}\sqrt{1-e^{2}})$
    \citep{bakos:2010:hat11}.
}
\tablenotetext{d}{
    Values for a quadratic law, adopted from the tabulations by
    \cite{claret:2004} according to the spectroscopic (SPC) parameters
    listed in \reftabl{stellar}.
}
\tablenotetext{e}{
    The 95\% confidence upper-limit on the eccentricity from a model
    in which the eccentricity is allowed to vary in the fit.
}
\tablenotetext{f}{
    Error term, either astrophysical or instrumental in origin, added
    in quadrature to the formal radial velocity errors. This term is
    varied in the fit assuming a prior inversely proportional to the
    jitter.
}
\tablenotetext{g}{
    Correlation coefficient between the planetary mass \mpl\ and
    radius \rpl\ determined from the parameter posterior distribution
    via $C(\mpl,\rpl) = <(\mpl - <\mpl>)(\rpl -
    <\rpl>)>/(\sigma_{\mpl}\sigma_{\rpl})>$ where $< \cdot >$ is the
    expectation value operator, and $\sigma_x$ is the standard
    deviation of parameter $x$.
}
\tablenotetext{h}{
    Planet equilibrium temperature averaged over the orbit, calculated
    assuming a Bond albedo of zero, and that flux is reradiated from
    the full planet surface.
}
\tablenotetext{i}{
    The Safronov number is given by $\Theta = \frac{1}{2}(V_{\rm
    esc}/V_{\rm orb})^2 = (a/\rpl)(\mpl / \mstar )$
    \citep[see][]{hansen:2007}.
}
\tablenotetext{j}{
    Incoming flux per unit surface area, averaged over the orbit.
}
\ifthenelse{\boolean{emulateapj}}{
  \end{deluxetable*}
}{
  \end{deluxetable}
}
%



\section{Discussion}
\label{sec:discussion}
\subsection{Density}
\label{sec:density}

\begin{figure}[ht]
\vspace{5mm}
\plotone{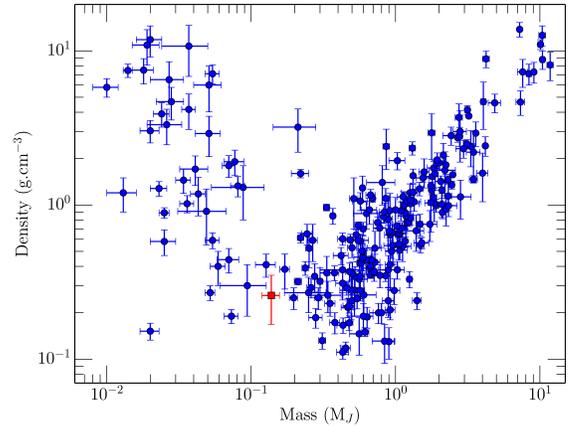}
\caption{
  The densities of exoplanets over four orders of mass. A
  characteristic ``V'' shape denotes two regimes of planets.  The low
  mass planets ($<\sim$0.1\mjup) show bulk densities that decrease
  with mass, as the gaseous planetary atmospheres grow in size
  relative to the solid fraction of the planets.  The high mass
  planets ($>\sim$0.1\mjup) as gas planets where density increases
  with mass due to partial electron degeneracy in the planet interior.
  \hatcurb{} lies in the currently sparsely sampled transition region
  between these two regimes.  All known exoplanets with well
  characterized densities (density uncertainties less than 40\%) are
  plotted as blue circles, with data from the NASA Exoplanet Archive
  \citep{akeson:2013:exoplanets} as of 2015 March 25. \hatcurb{} is
  plotted as red square.
\label{fig:density}}
\end{figure}

\hatcurb{} is a very low density exoplanet, with a mean density of
just \hatcurPPrho\,\gcmc.  Figure~\ref{fig:density} shows \hatcurb{}
in context with all known exoplanets with precisely measured
densities.  We see from Figure~\ref{fig:density} that \hatcurb{} sits
in a transition region: between the low-mass, non-degenerate planets
where bulk density decreases with total mass and the high mass,
partially degenerate gas giants where bulk density increases with
total mass.  Discovering transiting super-Neptunes in this transition
region is important, as it allows us to investigate the properties of
planets that have not undergone ``run-away'' gas-accretion
\citep{mordasini:2009}, in contrast to the well-studied population of
hot-Jupiters.

\hatcurb{} is closest in mass and density to the exoplanet HAT-P-18b
\citep{hartman:2011:hat1819}, which has a similar density (0.25 \gcmc)
but is slightly more massive at M=0.197\mjup.  Using the
\citet{fortney:2007} models for gas giant planets, and assuming an age
of the system to be approximately 4.5 Gyrs as indicated from the YY
isochrones (see Figure~\ref{fig:iso}), \hatcurb{} would have a
core-mass of just 10\mearth.  As with HAT-P-18, the metallicity of
\hatcur{} is not low (\feh=\hatcurSMEiizfeh), and therefore the
low-mass core that we infer is unlikely to be related to a lack of
metals in the proto-planetary disk from which \hatcurb{} formed.  This
argues against the metallicity-radius relationship for low-mass gas
giants as discussed in \citet{faedi:2011}.  Purely in terms of mass,
\hatcurb{} is similar to the exoplanet Kepler-101b, which is slightly
more massive at 0.16\mjup\ \citep{bonomo:2014:kepler101}.  However
Kepler-101b has a radius of just 0.515\rjup\ and a resulting bulk
density of 1.45\gcmc, which is 5.6 times higher than \hatcurb.  This
huge diversity in densities exhibited by super-Neptunes points to very
different formation scenarios for these systems.  In order to
understand these differences, however, it is incumbent on both precise
ground-based surveys and future space-based transit surveys to expand
the population of known super-Neptunes.

We note that \hatcurb{} has a mass and period that puts it right on
the edge of the `sub--Jupiter desert' proposed by \citet{szabo:2011}.
This desert may be due to evaporation \citep{kurokawa:2014}, in which
case \hatcurb{} is at the orbital limit for super-Neptunes, and if it
were any closer to its host star then evaporation may have reduced
it to a super-Earth class exoplanet.  The fact that \hatcur{} is
currently a quiet star (see Section~\ref{sec:hispec}) may also point
to a history of comparatively low X-ray flux, lowering the mean
evaporation rate of \hatcurb{} and permitting it to exist on the edge
of the ``sub-Jupiter desert''.


\subsection{Ground-based Sensitivity}
\label{sec:ground}
\begin{figure}[hb]
\vspace{5mm}
\plotone{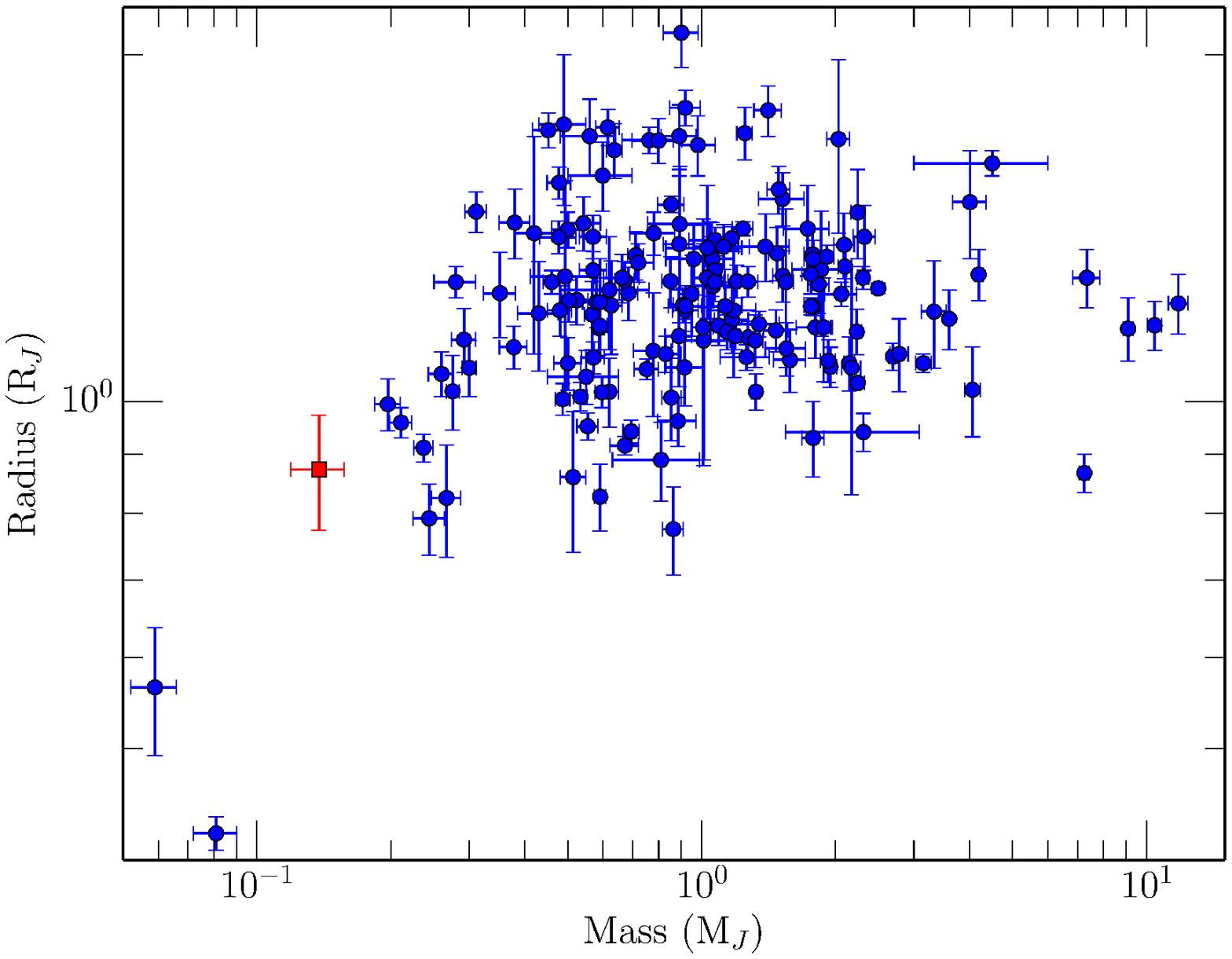}
\caption{
  Mass-radius diagram for all exoplanets discovered with wide-field
  transit surveys from the ground (WASP, HAT, TrES, XO, KELT, OGLE,
  Qatar, and WTS).  Data from the NASA Exoplanet Archive
  \citep{akeson:2013:exoplanets} as of 2015 March 25. \hatcurb{} is
  plotted as red square, and is the third lowest mass exoplanet
  discovered by wide-field, ground-based searches with a mass of just
  $\hatcurPPm$\,\mjup{}.
\label{fig:mass-radius}}
\end{figure}

While the period of \hatcurb{} is typical for an exoplanet detected by
a wide-field, ground-based transit survey (P=\hatcurLCP\,d), the
derived mass (M=\hatcurPPm \, \mjup) and to a lesser extent the radius
(R=\hatcurPPr \,\rjup), are much lower than typical, as is displayed
in Figure~\ref{fig:mass-radius}.  In fact \hatcurb{} is the third
lowest mass planet to be detected by a wide-field, ground-based
transit survey: only HAT-P-26b \citep{hartman:2011:hat26} and
HAT-P-11b \citep{bakos:2010:hat11} are less massive.  The discovery
demonstrates that the \hs{} global network is capable of discovering
gas planets in this low-mass regime.  The importance of Keck/HIRES in
this discovery should also be emphasized; given the magnitude of
\hatcur{} and the mass of \hatcurb{}, Keck/HIRES is one of the only
facilities in the world capable of confirming this discovery.  The
ability to use a synthetic template (see Section~\ref{sec:hispec}) for
cross-correlation is also a significant development for faint targets
such as \hatcur, for not only can it result in more precise radial
velocity measurements, but it can also potentially dispense with the
time-consuming step of creating an iodine-free template of the star.

It is interesting to consider whether \hatcurb{} and other transiting
exoplanets detected from ground-based transit surveys should be
included as targets on upcoming space missions such as TESS
\citep{ricker:2014:TESS} and PLATO \citep{rauer:2014:PLATO}.
Nominally \hatcur{} would fall outside the magnitude range of the TESS
mission ($I_{C}<12$ for F,G,K stars), and so it would only be able to
be monitored at a 30-minute cadence via the TESS full-framed images.
However there are reasonable arguments to make that it would be wise
to select the 4 pixels which \hatcur{} would occupy in order to obtain
2-minute cadence photometry.  First, a set of well-sampled transits
observed in 2019 would provide a long baseline with which to measure
if the orbit had evolved in terms of transit timing variations.
Secondly, the results from the Kepler and HARPS surveys for exoplanets
show that many planetary systems are very well aligned
\citep{figueira:2012}, making it much more probable that additional
planets could be discovered if we monitor a known transiting system.
Although most hot Jupiters do not appear to have close
planetary companions \citep{steffen:2012}, it is possible that
super-Neptunes such as \hatcurb{} would not follow this trend. Finally,
precise photometry may help us understand this system in more detail,
such as providing a stellar rotation rate or probing star spots if the
planet happens to occult an active region of the star during transits
\cite[e.g.][]{mohlerfischer:2013:hats2}.

\subsection{Neighbors}
\label{sec:neighbour}
\begin{figure} [hb]
\vspace{5mm}
\plotone{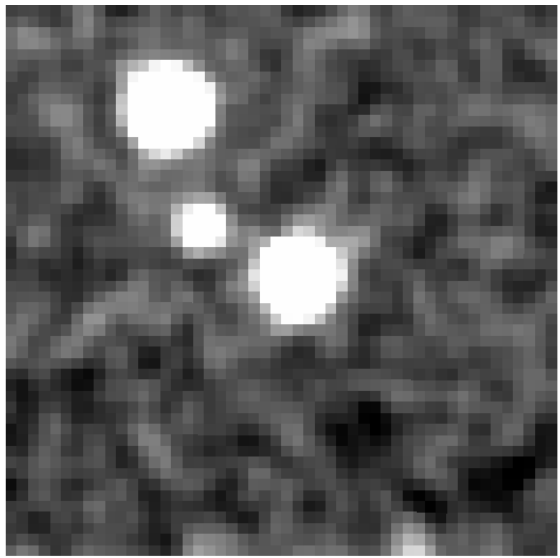}
\caption{
50\arcsec $\times$ 50\arcsec\ 2MASS \band{j} image centered on
\hatcur{} (north-up, east to the left).  The faint neighbor (2MASS
19394667-2544496) to the north-east is 10\arcsec from \hatcur{}, while
the brighter neighbor to the north-east (2MASS 19394689-2544386) is at
19.4\arcsec.  The bright neighbor is very similar in apparent
magnitude to \hatcur{} ($\delta$J+0.4mag), and is therefore an ideal
reference star for long-slit transit spectroscopy or narrow FOV,
high-precision transit photometry.  
\label{fig:neighbour}}
\end{figure}

\hatcur{} is in a relatively crowded field (see
Table~\ref{tab:stellar}), that lies approximately $20^{\circ}$ off the
Galactic plane.  While there is no evidence of a blended neighbor very
close ($<5$\arcsec) to \hatcur{}, there are two neighboring stars at
10\arcsec{} and 19.4\arcsec, which are displayed in
Figure~\ref{fig:neighbour}. At the estimated distance to \hatcur{}
(\hatcurXdistred\,pc), these neighboring stars are not physically
associated with \hatcur.  They also do not significantly affect the
\hs{} photometry, as the \hs{} pixel scale (3.7\arcsec\ per pixel)
means the neighbors are spatially resolved in the \hs{} images and in
our follow-up photometry.  However the neighbor at
19.4\arcsec\ (\bneighbor) is almost exactly the same magnitude (just
0.4\,mag brighter) and with a similar color: $(J-K)=0.512$ compared
with $(J-K)=0.437$ for \hatcur.  This makes \bneighbor{}\@ an ideal
reference star for long-slit, high resolution transit spectroscopy of
\hatcur.  Such observations have been carried out from ground-based
facilities in other systems where there is a nearby reference star
available \citep[e.g. XO-2b,][]{sing:2012}.  We also expect \hatcurb{}
to have an atmospheric scale height of 927\,km if we assume a Neptune
like mean molecular weight ($\mu$=2.53\,g.mol$^{-1}$).  Such a large scale
height would increase the strength of the absorption signal for
transmission spectroscopy.  The primary limitation will, of
course, be the fact that \hatcur{} is relatively faint at
V=\hatcurCCmag, so building up high signal-to-noise data will be difficult.

\acknowledgements 

\paragraph{Acknowledgements}
Development of the \hs{} project was funded by NSF MRI grant
NSF/AST-0723074, operations have been supported by NASA grants
NNX09AB29G / NNX12AH91H and internal Princeton funds. Follow-up
observations receive partial support from grant NSF/AST-1108686.
A.J.\ acknowledges support from FONDECYT project 1130857, BASAL CATA
PFB-06, and project IC120009 ``Millennium Institute of Astrophysics
(MAS)'' of the Millenium Science Initiative, Chilean Ministry of
Economy. R.B.\ and N.E.\ are supported by CONICYT-PCHA/Doctorado
Nacional. R.B.\ and N.E.\ acknowledge additional support from project
IC120009 ``Millenium Institute of Astrophysics (MAS)'' of the
Millennium Science Initiative, Chilean Ministry of Economy.
V.S.\ acknowledges support form BASAL CATA PFB-06.
K.P. acknowledges support from NASA grant NNX13AQ62G.
B.J.F acknowledges support from NSF Graduate Research Fellowship Grant
No. 2014184874.  Any opinion, findings, and conclusions or
recommendations expressed in this material are those of the authors
and do not necessarily reflect the views of the National Science
Foundation.
Operations at the MPG~2.2\,m Telescope are jointly performed by the
Max Planck Gesellschaft and the European Southern Observatory.
This work is based on observations made with ESO Telescopes at the La
Silla Observatory.
Observations from the duPont and Swope telescopes were taken as part
of programs CN2012A-61, CN2013A-171 and CN2014A-104 awarded by the
Chilean Telescope Allocation Committee (CNTAC).
This paper also uses observations obtained with facilities of the Las
Cumbres Observatory Global Telescope.
The radial velocity data presented herein were obtained at the
W.M. Keck Observatory, which is operated as a scientific partnership
among the California Institute of Technology, the University of
California and the National Aeronautics and Space Administration. The
Observatory was made possible by the generous financial support of the
W.M. Keck Foundation.  The authors wish to recognize and acknowledge
the very significant cultural role and reverence that the summit of
Mauna Kea has always had within the indigenous Hawaiian community.
Work at the Australian National University is supported by ARC Laureate
Fellowship Grant FL0992131.
This research has made use of the NASA Exoplanet Archive, which is
operated by the California Institute of Technology, under contract
with the National Aeronautics and Space Administration under the
Exoplanet Exploration Program.


\end{document}